\documentclass[pra,amssymb,amsfonts,twocolumn]{revtex4}
\usepackage{epsf,latexsym}
\usepackage{amsmath}
\usepackage[mathscr]{eucal}

\usepackage{times,graphics}
\usepackage[final]{pdfpages}
\usepackage{amscd}
\usepackage{epsfig}

  \def\be{\begin{equation}}
  \def\ee{\end{equation}}
  \def\bea{\begin{eqnarray}}
\def\eea{\end{eqnarray}}
  \def\6{\langle}
  \def\9{\rangle}
  \def\pad{\partial}
  \def\tr{{\rm{tr}}}
  \def\half{\mbox{$1\over2$}}
  \def\cH{{\cal H}}

    \def\bof{{\bf f}}
    \def\bk{{\bf k}}
  \def\bj{{\bf j}}
  \def\bp{{\bf p}}
  \def\bq{{\bf q}}
  \def\bx{{\bf x}}
   \def\by{{\bf y}}

 \def\bA{{\bf A}}
     \def\bB{{\bf B}}
        \def\bE{{\bf E}}
\def\bF{{\bf F}}

\def\bep{\mbox{\boldmath $\epsilon$}}
\def\bphi{\mbox{\boldmath $\phi$}}

\def\hphi{{\hat{\phi}}}

\def\hA{{\hat{A}}}
\def\hB{{\hat{B}}}
\def\hH{{\hat{H}}}
\def\hE{{\hat{E}}}

\def\hQ{{\hat{Q}}}
\def\hT{{\hat{T}}}
\def\hU{{\hat{U}}}
\def\hX{{\hat{X}}}

\def\hPi{{\hat{\Pi}}}

\def\hbB{{\hat{\bB}}}
\def\hbE{{\hat{\bE}}}
\def\hbF{{\hat{\bF}}}
\def\hbN{{\hat{\bold{N}}}}

\def\mR{{\mathbb{R}}}

\begin{document}

\title{Localization of relativistic particles and uncertainty relations}

\author{Daniel R. Terno}
\affiliation{Department of Physics \& Astronomy, Macquarie University, Sydney, NSW 2109,  Australia}

\begin{abstract}
Localization of relativistic particles and their position-momentum uncertainty relations are  not yet fully understood. We discuss two schemes of photon localization that are based on the energy density. One scheme produces a positive operator-valued measure for localization. It   coincides with the number density operator and   reproduces an effective $3\times 3$ polarization density matrix. Another scheme results in a probability distribution that is conditioned on the detection. In both schemes   the uncertainty relations for transversal position and momentum approach the Heisenberg bound $\Delta p \Delta z=\half\hbar$. 
\end{abstract}
\pacs{03.67.-a,  11.10.-z, 42.50.-p}
\maketitle

\section{Introduction}

Localized detection events provide the experimental basis of quantum field theory and quantum optics \cite{ha,qft:b,mw95}. They are critical for   understanding   relativistic aspects of quantum information processing \cite{pt04,rqi}.
Nevertheless,   analysis  of localization in the relativistic regime \cite{lan,bi35,nw49,wi62,mandel,amrein,schl71,hg74,heg85,
wer86,ali98,toller,bu99,bb96,kel05,bb2-12a,hans1} is still incomplete.
A stark contrast between our ability to
 manipulate single photons \cite{manypho,manypho1} or trace molecules with   subwavelength  accuracy \cite{nano},  and the lack of clarity in describing the photon localization  is illustrated by a recent controversy \cite{bb2-12b,com12}.

One can consider localization in a space-time or on a given time slice. The solution to the former problem is given by the probability of detecting a particle   during an interval of time $\Delta t$ around time $t$ in a volume $\Delta V$ around $\bx$. If the detection time $\Delta t$ can be ignored (see \cite{peres00}), the problem reduces to describing   localization on a time slice $t$. Through the paper we consider a flat space-time, use the $+---$ signature of the metric, and  set $\hbar=c=1$, unless specified otherwise. We focus on single-particle states.

Complexity of relativistic localization is highlighted by the absence of
a self-adjoint time operator \cite{pauli},
and thus absence of space-time localization  operators $\hX^\mu$, $\mu=0,\ldots, 3$ \cite{wi62,toller} (For   recent discussions see \cite{timeop}). There is
no unique spatial position operator, and (depending on the imposed  requirements) there is no position
operator for massless particles with
spin larger than $\half$ \cite{nw49,wi62}. When the operator exists its components may be non-commutative,
$[\hX^k,\hX^j]\neq0$ \cite{schw1}, and
the resulting probability distribution may lack a probability current or have causality problems
\cite{wi62,nw49,schl71,hg74,heg85,ali98}.

 The
Newton--Wigner wave function $\psi^{NW}$  \cite{nw49,ha}  is a popular tool to describe localization of
massive particles, but it is only a partial remedy \cite{ali98, bk01}. If $|\psi^{NW}(t,\bx)|^2$ is taken to be a
  spatial probability distribution  then there are states that violate causality
  \cite{ru87} in the spirit of the Hegerfeldt's theorem \cite{heg85} (Sec.~III).
   In the framework of the algebraic field theory \cite{ha,qft:b} it was shown that the localization description
 cannot be realized by local  or   quasi-local operators \cite{toller,gian}.

Absence of   \textit{a priori} preferable descriptions of localization  motivates an operational approach. A way to substantiate the
notion of particles and to reconcile it with the local  quantum field theory is to analyze  the (model)
detectors' excitations.  Investigating the space-time localization one considers a  response rate of such a  detector   \cite{bd,ueff,dete1}, which leads to a Lorentz-invariant detection rate $w(t,\bx|\rho)\equiv w(x|\rho)$, where the detection probability for a state $\rho$ in the space-time four-volume $dtdV$ near the space-time point $x$ is
\be
d P(x,d\Omega|\rho)=w(x|\rho)dtdV
\ee
Localization on a time-slice    selects a Lorentz frame that is associated with the detector(s). In this case the quantities of interest are the probability density $p(t,\bx|\rho)$ and its current. Their interpretation   as describing the evolution of the particle's position in time is not straightforward. In particular, unavoidable dark counts (response of a local detector to vacuum) that give ``false positives"  and a finite probability of failure, when the detectors do not ``fire" even if a particle is present, should be taken into account.

In this article we focus on   localizing single-particle states on a time-slice.  The associated probability density is constructed using three ingredients:
the energy density of a quantized field, the particle's  momentum-space wave function, and its energy. Two possible solutions satisfy
the constraints that are imposed by quantum field theory, fulfill many of the desired properties suggested by
non-relativistic quantum mechanics, and relate to familiar objects in quantum optics.  We apply these constructions  to  the photonic uncertainty relations and discuss possible experiments.

The rest of the paper is organized as follows. We review the general properties of localization   in the next section. Section III deals with   properties of energy density and two possible methods of constructing normalized probability distributions. This general theory is applied to photons in Sec. IV, and the uncertainty relations are derived in Sec. V. Sec. VI discusses the implications of our results, their limitations and future directions.
\vspace{-2mm}

\section{General properties of localization}
\vspace{-2mm}

Mathematical difficulties with   orthogonal projections and detection analysis led to discussion of localization in terms of   positive
operator-valued measures (POVMs) \cite{ali98,bu99}.
  A POVM constitutes a non-orthogonal
decomposition of the
identity by means of positive operators  $\hPi(x)$, resulting in detection probabilities $P(x)=\tr\rho\hPi(x)$  for the events $\{x\}$ \cite{hol,povm}.

In a given Minkowski frame  the sets $\Delta\subset\mR^3$ correspond to the statements that the system is located in $\Delta$ at time $t$.
Each set is associated with a positive operator $\hPi(\Delta)$. The operators are not assumed to be local or quasilocal.
The operators form a decomposition of identity, $\hPi(\mR^3)=1$. For the disjoint sets $\Delta_1$ and $\Delta_2$ the resulting operator satisfies
\be \vspace{-0.5mm}
\hPi(\Delta_1\!\cup\!\Delta_2)=\hPi(\Delta_1)+\hPi(\Delta_2).\vspace{-0.5mm}
\ee

Behaviour under rotations and translations is the hallmark of a position observable. Let $g\!\cdot\!\Delta=R\Delta+\mathbf{b}$ denote  the action of the Euclidean group (the rotation $R$ followed by the translation $\mathbf{b}$) on $\Delta$, and by$\hU(g)$ its unitary representation. Then
\be
\vspace{-0.5mm}\hPi(g\!\cdot\!\Delta)=U(g)\hPi(\Delta)U(g)^\dag.\vspace{-0.5mm}
\ee

Several intuitively attractive features of position will not be realized. We do not require a sharp localizability \cite{bu99}: for the disjoint $\Delta_1$ and $\Delta_2$ the product $\hPi(\Delta_1)\hPi(\Delta_2)\neq 0$. Instead of position operators $\hX^l$ we have the first statistical moment operators,
\be
\hQ^l(t)=\int\!d^3\bx x^l\hPi(t,\bx), \qquad l=x,y,z,\vspace{-0.5mm}
\ee
where   we \! abuse  the notation  by   denoting \! the   measure   $\hPi(d^3\! \bx)$,
$$
\hPi(\Delta)=\int_\Delta \hPi(d^3\bx),
$$
as $\hPi(\bx)d^3\bx$.
The operators $\hQ^l$  do not commute. However, since the measurement is described by a single measure $\hPi$m they are evaluated jointly,
\be
\6\hQ^l(t)\9=\int\!d^3\bx x^l\,p(t,\bx), \qquad p(t,\bx)=\tr[\,\rho \hPi(t,\bx)],
\ee
where $\rho$ is density matrix of the system.

If the   probability distribution  $p(t,\bx)$ is interpreted as a time-dependent indicator of the approximate position of a particle, it is reasonable to expect the conserved current,
\be
\frac{\pad p (t,\bx)}{\pad t}=-\nabla\bj(t,\bx), \label{current1}
\ee
and a causal evolution of $p(t,\bx)$. Probability density should propagate causally. Both requirements are realized below.

The localization scheme should be Lorentz covariant. However, there are different levels at which it can be realized. Consider an inhomogeneous Lorentz transformation $x\rightarrow \Lambda x+ a$, where $a$ is a four-dimensional translation and $\Lambda$ some proper Lorentz transformations that connects two reference frames. For a POVM that results in the detection rate
\be
w(x|\rho)=\tr \rho \hPi_w(x),
\ee
the invariance of probabilities requires a unitary transformation of the POVM operators, and the state transformation law
\be
|\Psi\9\rightarrow\hU(a,\Lambda)|\Psi\9,
\ee
\vspace{-1mm} implies \cite{peres00}\vspace{-0.5mm}
\be
  \hPi_w(x)\rightarrow\hPi'_w(x')=\hU(a,\Lambda)\hPi_w(x)\hU(a,\Lambda)^\dag.\vspace{-0.5mm}
\ee
The constant time slice in one frame does not transform into a constant time slice in another frame, so no such manifest covariance can be required for $\hPi(t,\bx)$. Instead, we expect it to be constructed in the same way in both frames, where the constituent operators transform according to their appropriate laws. The probability conservation Eq.~\eqref{current1} should hold true in any frame. However, there is no reason to demand that $(\rho,\bj)$ form a four-vector and transform accordingly.

\section{Energy density and detection probability}
\vspace{-1mm}
\subsection{Properties}
\vspace{-1mm}
Energy density broadly agrees with our intuition of ``where  the particle is." Moreover,\! it propagates causally \cite{bk01} and
is directly related to photodetection
 \cite{mw95,bb96,kel05,amrein}. If the electrons in a detector
interact with the electric field of light, then a leading order detection probability is proportional to the
expectation value of the normal-ordered electric-field intensity
operator  \cite{mw95}, \!and  the \!latter  is \!proportional  to  the energy \!\! density.

In this section we discuss the resulting normalized probability distributions. We use a real scalar field for simplicity.
Consider  a one-particle state,
 \be
  |\Psi\9=\int d\mu(p)\psi(\bp)|\bp\9,\qquad
  d\mu(p)=\frac{d^3\bp}{(2\pi)^32p^0}, \label{scapsi}
  \ee
where $p^0=E_\bp=\sqrt{m^2+\bp^2}$,
$\6\bp|\bq\9=(2\pi)^3(2p^0)\delta(\bp-\bq)$, and $\int d\mu(p)
|\psi|^2 =1$.
   The inner product of two states is
  calculated  in  the momentum
  representation,
  \be
  \6\Psi|\Phi\9=\int d\mu(p)\psi^*(\bp)\phi(\bp).
  \ee
\vspace{-1.5mm}

 For a state
$|\Psi\9$ the energy density equals
\begin{align}
& T_{00}(t,\bx)  =  \6\Psi|{\hat{T}}_{00}(t,\bx)|\Psi\9 \nonumber \\
&=
|\nabla\psi(t,\bx)|^2+|\pad_t\psi(t,\bx)|^2+m^2|\psi(t,\bx)|^2,\label{dens}
\end{align}
where $\hT_{00}$ is the normal-ordered Hamiltonian density, and $\psi(t,\bx)$ is the configuration space wave function
(Appendix A). The energy density is positive and the
Lorentz transformation properties are built into this quantity by
 definition.

   Let us consider possible
causality violations. The Hegerfeldt's theorem in its strongest
version proves a superluminal speed for an exponentially localized
particle. If the probability of finding it outside a sphere of
radius $R$ is bounded by
\be
{\rm Prob}_{\not\in R}< C^2\exp(-2\gamma R),
\ee
where $C$ is some constant and $\gamma>m$, then the state will
spread faster than light. However, it was shown in \cite{bk01} that no physical state can satisfy this bound.
If $T_{00}(t,\bx)$ satisfies it, then both
  $|\psi(t,\bx)|$ and $|\pad_t\psi(t,\bx)|$ are bounded by $C\exp(-\gamma
R)$. It implies that both $\psi(\bp)$ and $\psi(\bp)/E_\bp$ are analytic
functions in the strip of the complex plane that is bounded by at least $|{\rm
Im}(\bp)|\leq m$, which is inconsistent with the branch cuts in $E_\bp$ at
$|\bp|=\pm im$. Therefore the energy density cannot be ``localized" enough to
violate
  causality.

\vspace{-2mm}

\subsection{Normalization}
\vspace{-2mm}

If the probability to detect  a particle in some volume $\Delta$ is proportional to the integral of the energy density over this volume, then the probability that a detection happened somewhere is
\be
P(\mR^3|\Psi)=K\int T_{00}(t,\bx)d^3\bx\equiv K\6\Psi|{\hH}|\Psi\9<1, \label{detpro}
\ee
\nopagebreak for some constant $K$.
As a result,  one way to obtain a  proba-\newpage \noindent bility   density for the particle's position is   rescale the  detection probability,
  \be \vspace{-0.5mm}
  P^E(\Delta|\Psi)=\int\Delta  \frac{T_{00}(t,\bx)}{{\6\Psi|{\hH}|\Psi\9}}d^3\bx, \qquad 0\leq P(\Delta|\Psi)\leq 1.
  \label{prob}
  \ee
In other words, this is a probability to find a particle in a region $\Delta$ given that a detection occurred at all.
 In contrast with the POVM outcomes \cite{pt98,pt04},  this quantity is not convex \cite{dt02}.  Consider a state
$\rho$ which is a mixture of  two pure states $|\Psi_1\9$ and
$|\Psi_2\9$,
  \be
  \rho=a\rho_1+(1-a)\rho_2, \qquad 0<a<1.
  \ee
According to Eq.~(\ref{prob}) probability densities for the states
$\rho_i=|\Psi_i\9\6\Psi_i|$ are
  \be
  p_i^E(t,\bx)=
 \frac{\tr\,\rho_i \hT_{00}(t,\bx)}{\tr\,\rho_i \hH}.
  \ee
 By using either the above definition or the convexity property we reach \vspace{-1mm}
\begin{align}
  p_\rho^E(t,\bx) &  =  \frac{\tr\,\rho\hT_{00}(t,\bx)}{\tr\,\rho \hH} \nonumber \\
  & \neq  \frac{a\tr\,\rho_1 \hT_{00}(t,\bx)+(1-a)\tr\,\rho_2 \hT_{00}(t,\bx)}
  {a\tr\,\rho_1 \hH +(1-a)\tr\,\rho_2 \hH},\label{contra}
  \end{align}
and these two expressions are generally different. This behaviour is typical for post-selected quantities \cite{t99}.

Linearity is restored if  normalization of the probability distribution follows   the operator-valued measure.  We  construct a POVM element as
  \be
  {\hPi}(t,\bx)={\hH}^{-1/2}{\hT_{00}}(t,\bx){\hH}^{-1/2}.\label{povm}
  \ee
The action of $\hH^{-1}$ is   is well
defined when we restrict it to the non-vacuum states. It was shown in \cite{bb2-12a}
that the Born-Infeld position operator \cite{bi35} that is obtained from the
first moment of the energy distribution $\hbN=\int d^3\bx\, \bx \hT_{00}$ equals   the first moment  of this operator density,
\vspace{-1mm}
\be
\half(\hH^{-1}\hbN+\hbN \hH^{-1})=\int\! d^3\bx\,\bx \hPi(t,\bx). \label{xbb}
\ee

\vspace{-2mm}

The probability density $\6\Psi|\hPi(t,\bx)|\Psi\9$
can be written as the energy density of a classical field,
\begin{align}
& p(t,\bx) :=  \6\Psi|{\hPi}(t,\bx)|\Psi\9 \nonumber \\
&~~~~= \big(|\nabla\tilde{\psi}(t,\bx)|^2+|\pad_t\tilde{\psi}(t,\bx)|^2
  +m^2|\tilde{\psi}(t,\bx)|^2 \big)\label{newp},
  \end{align}
  where the additional $E^{-1/2}_\bp$ factor  is added to $\psi(\bp)$,
  \be
  \tilde{\psi}(t,\bx)=\int
  d\mu(p)\frac{\tilde{\psi}}{\sqrt{E_\bp}}(\bp)e^{i(\bp\cdot\bx-Et)}.
  \ee
The probability current $j^{\,l}(t,\bx)=\6\hT^{0l}\9$ is  obtained similarly (Appendix A). Working in the
momentum space representation it is easy to see that
that indeed
  \be
 p\geq 0, \qquad \int p(t,\bx)d^3\bx=1, \qquad \dot{p}=-\nabla\cdot\bj,
  \ee
 where  the energy-momentum conservation $\pad_\mu\hT^{0\mu}=0$ results in the continuity equation for the probability density.

   The arguments of \cite{bk01}
   for the energy density are equally well applied to our probability
  distribution $p$, so Hegerfeldt's theorem does not present a paradox. We note that presence
  of the Hamiltonian   makes $\hPi$ a non-local   operator, in agrement with \cite{toller, gian}.
  A non-relativistic limit is obtained by expanding $p(t,\bx)$ in powers of $\bp/m$ (Appendix A).

\vspace{-1mm}
\section{Localization of photons}
\vspace{-2mm}
 Following \cite{bb2-12a,bb2-12b} we use the Riemann-Silberstein vector
\be
\bF:=\bE+i\bB. 
\ee
  The classical energy density in a free space is then
\be
T_{00}=\half(\bE^2+\bB^2)=\half\bF^*\!\cdot\!\bF
\ee
We work in the radiation gauge (so $\bE=-\dot{\bA}$ and $\bB=\nabla\times\bA$) and write the  solutions of the wave equation
with the help of right and left
polarization vectors $\bep_\bp^\pm$,
\be
\bp \times \bep^\pm_\bp=\mp i|\bp|\bep^\pm_\bp,
\ee
that   satisfy the   convention
\be
\bep_\bp:=\bep^+_\bp=\bep^{-*}_\bp=\bep_{-\bp}^*, \qquad \bep^*_\bp\cdot\bep_\bp=1, \quad \bep_\bp^2=0.
\ee
We adapt   Wigner's construction of the massless representation of the Poincar\'{e} group to the construction of the polarization
basis vectors $\bep_\bp^\pm$ \cite{pt04,lpt03}. For $\bp=p(\sin\theta\cos\varphi,\sin\theta\sin\varphi,\cos\theta)$
\be
\bep_\bp=\tfrac{1}{\sqrt{2}}\big(\cos\theta\cos\varphi-i\sin\varphi, \cos\theta\sin\varphi+i\cos\varphi, -\sin\theta\big).
\ee

Writing the quantum field as
\be
\hat{\bA}(t,\bx)=\int\! d\mu(p)\sum_{\lambda=\pm 1}\left(\bep^\lambda_\bp\alpha_{\bp\lambda}e^{-ip\cdot x}+
\bep^{\lambda*}_\bp\alpha_{\bp\lambda}^\dag e^{ip\cdot x}\right),
\ee
we obtain the standard commutation relations (Appendix B), and
the Riemann-Silberstein operator
\be
\hbF=\hbE+i\hbB=2i\int\! d\mu(p) E_\bp \bep _\bp\big(\alpha_{\bp+}e^{-ip\cdot x }+
 \alpha_{\bp-}^\dag e^{ip\cdot x } \big).
\ee
Energy density then becomes
\begin{align}
 \hT_{00} & =\frac{1}{2}::\hbF^\dag\!\cdot\!\hbF::  \\
&=2\sum_{\lambda=\pm 1}\int\!d\mu(p)d\mu(q)E_\bp E_\bq\,
 \bep_\bp^*\!\cdot\!\bep_\bq\alpha^\dag_{\bp\lambda}\alpha_{\bq\lambda}
 e^{i(p-q)\cdot x }, \nonumber
\end{align}
where the terms that annihilate one-particle  states are dropped.
 A straightforward calculation leads to a POVM
\begin{align}
&\hPi(t,\bx) =\hat{H}^{-1/2}\hT_{00}(t,\bx)\hat{H}^{-1/2} \label{povm-pho} \\
 &=2\sum_{\lambda=\pm 1}\int\!d\mu(p)d\mu(q)\sqrt{E_\bp E_\bq}
 \bep_\bp^*\!\cdot\!\bep_\bq\alpha^\dag_{\bp\lambda}\alpha_{\bq\lambda}
 e^{i(p-q)\cdot x }. \nonumber
\end{align}
For a generic one-photon state
\be
|\Psi\9=\sum_{\lambda=\pm 1}\!\int d\mu(p)\psi_\lambda(\bp)a^\dag_{\bp\lambda}|0\9,
\ee
the energy density can be compactly written  with the  3-vector wave function \cite{mw95}
\be
\bphi_\lambda(t,\bx)=\frac{1}{(2\pi)^3}\int\!d^3\bp  \bep_\bp^\lambda \psi_\lambda(\bp)e^{-i p\cdot x },
\ee
as
\be
T_{00}(x)=\6\Psi|\hT_{00}(x)|\Psi\9=\frac{1}{2}\sum_{\lambda=\pm 1}|\bphi_\lambda(x)|^2.
\ee
On the other hand, the probability density is given by the Newton-Wigner wave function,
\be
p(t,\bx)=\6\Psi|\hPi(t,\bx)|\Psi\9=\sum_{\lambda=\pm 1}|\mathbf{f}_\lambda(t,\bx)|^2, \label{probph}
\ee
where
\be
\mathbf{f}_\lambda(t,\bx) =
\int\!\frac{d^3\bp}{(2\pi)^3}\bof_\lambda(\bp)e^{-i p\cdot x }
=\int\!\frac{d^3\bp}{(2\pi)^3} \bep_\bp^\lambda
\frac{\psi_\lambda(\bp)}{\sqrt{2E_\bp}}e^{-i p\cdot x }.\label{nw-back}
\ee

Several observations are in order. The configuration space wave function has the Newton-Wigner form  (see, e.g., \cite{mh12}), even if it is not an eigenfunction of the position operator. The POVM $\hPi$ is identical with Mandel's number density operator \cite{mandel}. The two wave functions are related as
\be
\bphi_\lambda(\bp)=\sqrt{2E_\bp}\,\mathbf{f}_{\lambda}(\bp).\label{conmom}
\ee
Effective polarization density matrix was introduced on the formal grounds \cite{pt04,lpt03} as an analog of the
 3$\times$3 classical correlation matrix \cite{mw95}. It results in the  probabilities $P_a(\rho)=\tr\, \Pi_a\rho$ for a class
of POVMs $\{\Pi_a\}$ that are based on a simple photodetection model \cite{lt05}. This density matrix is given by
\cite{pt04,lt05}
\be
\rho_{mn}=\int d\mu(p)\sum_{\lambda,\lambda'=
\pm 1}\psi_\lambda(\bp)\psi_{\lambda'}(\bp)^*\big(\bep^\lambda_{\bp}\big)_m\big(\bep^{\lambda'}_{\bp}\big)^*_n.
\ee
Introducing the vectorial components of the wave function as $f_m=(\bof_+)_m+(\bof_-)_m$, we see from Eq.~\eqref{nw-back} that
\be
\rho_{mn}=\int\!\frac{d^3\bp}{(2\pi)^3}f_m(\bp)f_n(\bp)^*=\int\!d^3\bx f_m(t,\bx)f_n(t,\bx)^*,
\ee
again similarly to the non-relativistic quantum mechanics.

\vspace{-1mm}
\section{Uncertainty relations}
\vspace{-2mm}
The uncertainty relations that involve  the position operator  were analyzed by
Bialynicki-Birula and  Bialynicka-Birula in \cite{bb2-12b}. Here instead we consider   statistics derived from the
probability density of Eq.~\eqref{probph} and  show that the product of variances
 approaches the Heisenberg bound $\Delta q \Delta p \geq \hbar/2$. Consider an one-photon state
with Gaussian profiles in both $x$  (the average propagation direction) and the transversal ($y$ and $z$)
components of the momentum, which is
built as a superposition of the right-polarized components only,
\begin{align}
f(\bp)&=\frac{\psi_+(\bp)}{\sqrt{2E_\bp}}  \label{exsat} \\
&=\frac{{2}{\pi}^{\tiny{3/4}}{w}} {\sqrt{\sigma}}
\exp\left(-\frac{w^2 (p_y^2+p_z^2)}{4}\right)
\exp\left(-\frac{(p_x-p_0)^2}{2\sigma^2}\right). \nonumber
\end{align}
For a classical beam $w$ is the
radius at which the intensity falls off to $1/2e^2$ of its maximum value on the axis of symmetry \cite{mw95,nano}, and
we take $1/w,\sigma\ll p_0$. Having in mind the actual experimental set-ups we   consider the transversal uncertainty relations.

Momentum space wave function has the standard probabilistic interpretation, i.e., for any power $k$
\be
\6p_l^{~k}\9=\int d\mu(p)p_l^{~k}|\psi_+(\bp)|^2=\int d^3\bp \,p_l^{~k}|f(\bp)|^2,
\ee
where $l=x,y,z$.  The first moments are $\6p_y\9=\6p_z\9=0$, $\6p_x\9=p_0$, and
the second moments of the transversal momentum are
\be
\6p_y^2\9=\6p_z^2\9=1/w^2. \label{momun}
\ee

 We expand the polarization vector $\bep_\bp$ in Eq.~\eqref{nw-back} in the inverse powers of $p_0$ and perform the Gaussian integrations, obtaining
\be
p(t,\bx)=p_0(t,\bx)\left(1+\frac{4(y^2+z^2)-2w^2}{p_0^2w^4}+{\cal{O}}(p_0^{-4})\right),
\ee
where $p_0(t,\bx)=|f(t,\bx)|^2$ is the scalar Gaussian probability density. As a result,
 $\6y\9=\6z\9=0$, and
\be
\6z^2\9=\6 y^2\9 =\frac{w^2}{4}+\frac{1}{2p_0^2}+{\cal{O}}(p_0^{-4}),
\ee
with the product of variances approaching  the usual non-relativistic bound
\be
\Delta z\Delta p_z\geq\frac{\hbar}{2}\left(1+\frac{\hbar^2}{p_0^2w^2}+{\cal{O}}(p_0^{-4})\right),
\ee
with the identical expression for the $y$ component, and we restored the use of $\hbar$.

Interpretation of $\mathbf{f}_\lambda(t,\bx)$  as the position probability amplitude faces a difficulty when contrasted with  the detection probability of Eq.~\eqref{detpro}. Because of Eq.~\eqref{conmom} the photon that is ``localized" in a bounded region $\Delta$ (i.e., $\mathbf{f}_\lambda(t,\bx)=0$ for$\bx\in\mR^3\!\setminus \!\Delta$) has a non-zero detection probability outside it \cite{mw95,amrein}. Therefore, it is useful to estimate the uncertainty using the normalized counting statistics that is represented by
 \be
 p^E(t,\bx)=\frac{|\bphi_+(t,\bx)|^2+|\bphi_-(t,\bx)|^2}{2E},
 \ee
where the expectation of energy is
\be
\qquad E=\half\int\!d^3\bx\big(|\bphi_+(t,\bx)|^2+|\bphi_-(t,\bx)|^2\big).
\ee
Expanding
\be
\sqrt{E_\bp}=\big([p_0+(p_x-p_0)]^2+p_y^2+p_z^2\big)^{1/4}
\ee
in the inverse powers of the momentum $p_0$ allows to obtain
 the series expansion of $\bphi(t,\bx)$  and $E$.   Statistical moments can be obtained either from the resulting probability density $p^E$ or by the techniques of Appendix A. The uncertainty  for the state $\psi_+$ of Eq.~\eqref{exsat} becomes
 \begin{align}
\6 y^2\9_E & =\6 z^2\9_E=\frac{1}{4}\left(w^2+\frac{1}{p_0^2}+{\cal{O}}(p_0^{-4})\right) \nonumber \\
&=\frac{1}{4}\left(w^2+\frac{\lambda_0^2}{4\pi^2}+{\cal{O}}(\lambda_0^{4})\right),
 \label{z2}
 \end{align}
 where $\lambda_0=h\!/\!p_0$ is the peak wavelength. Transversal momentum measurements are performed with he same detectors that are  positioned behind a suitably arranged lens system \cite{manypho,tas}. The latter implements  the Fourier transform between $\bphi(\bx)$  and $\bphi(\bp)$. Using $\bphi(\bp)$ to calculate the statistical moments leads to
\begin{align}
\6 p_z^2\9_E &=\6 p_y^2\9_E=\frac{1}{w^2}\left(1+\frac{1}{p_0^2w^2}+{\cal{O}}(p_0^{-4})\right) \nonumber \\
& =\frac{1}{w^2}\left(1+\frac{\lambda_0^2}{4\pi^2w^2}+{\cal{O}}(\lambda_0^{4})\right).
\end{align}

 \section{Discussion}

 Within the domain of validly of a single-particle picture the
energy density provides a satisfactory description of  localization.  By dropping the insistence
on local operators,
we not only conform to the general results of the quantum field theory, but break  a vicious cycle where the only operator
that satisfies all the ``reasonable" requirements of localization is identically zero.

For photons the resulting POVM $\hPi(t,\bx)$ coincides with the photon density operator. The resulting probability density shows a curious property: it is given by the
absolute value squared of the configuration space wave function of the Newton-Wigner form, even if the
Newton-Wigner position operator for photons does not exist.
The effective density matrix is a necessary mode of polarization
description  if the wave packet spread is important. Our scheme provides a direct
 link between the momentum space and configuration space descriptions of polarization.

We can settle the controversy about the
 photon uncertainty relations. Our scheme results in the expressions for probability density that are
  identical with that of \cite{bb2-12b}, but by
 accepting the standard definitions of the statistical analysis approach the non-relativistic
 Heisenberg bound $\Delta q\Delta p=\hbar/2$. It is remains to be seen how close it is possible to reach this limit by optimizing the state $\psi$.

The one-particle picture has a limited validity. Particles
  cannot be confined into a volume with a typical dimension  smaller than
  \be
  \sqrt[3]\Delta> \frac{1}{\6E\9},
  \ee
  \vspace{1mm}
   where $\6E\9$ is the particle's expected energy \cite{lan}. The very concept of
particles takes somewhat a `nebulous character' \cite{bd} in
field theories in a curved space-time or for accelerated observers.
A simple example of when this construction is not applicable is the Unruh effect \cite{ueff}.
An accelerated detector that moves in the Minkowski
vacuum responds as an inertial detector would if immersed into a
thermal bath of the temperature
\be
  T=a/(2\pi k_B),
\ee
where $k_B$ is Boltzmann's constant and $a$ the proper
acceleration. 
However,
the expectation of the renormalized stress-energy tensor is zero
in {both} inertial and accelerated frames.

  In   more complicated settings the question of positivity becomes acute.
Classical energy density is always positive, which is to say that
the stress-energy tensor for a scalar field satisfies the weak
energy condition (WEC)
  $T_{\mu\nu}u^\mu u^\nu\geq 0$, where $u^\mu$ is a causal vector.
In  quantum field theory \cite{ha,qft:b} it is
impossible. There are states $|\Upsilon\9$ that violate WEC, namely
$\6\Upsilon|{T}_{\mu\nu}u^\mu u^\nu|\Upsilon\9 \leq 0$ holds
\cite{nc65}, where $T_{\mu\nu}$ now is a renormalized stress energy operator.
 For example, squeezed states of electromagnetic
\cite{mw95} or scalar field have negative energy densities \cite{fr13,sq,f01}.
  It is known that even if  WEC is violated  the average WEC
still holds when the averaging is done over the world line of a
geodesic observer (inertial observer in the Minkowski spacetime)
\cite{tip}. There are also more stringent quantum inequalities
that limit the amount of the WEC violation. Instead of an infinite
time interval they deal with a sampling that is described by a
function with a typical width $t_0$ \cite{f01}.  The behaviour of
fields  subjected to boundary conditions is more complicated, but
similar constraints exist also in these cases
\cite{f01}.
   To meet our
ends we need the analogous inequalities to hold for a spatial averaging. This
is, however, impossible. A class of quantum states was constructed for a
massless, minimally coupled free scalar field  (superposition of the vacuum and
multi-mode two-particle states). These states can produce an arbitrarily large
amount of negative energy in a given finite region of space at a fixed time
 \cite{f02}. In this and similar cases the  spatial averaging over
  part of a constant time surface does not produce a positive quantity.
 An interesting line of research is to trace the emergence of well-defined particles in these complicated situations.

\acknowledgments

It is a pleasure to thank Iwo Bialynicki-Birula, Zofia Bialynicka-Birula, Lucas C\'{e}leri, Viktor Dodonov,
 Netanel Lindner,  Asher Peres, Federico Piazza, Paulo Souto Ribeiro and Hans Westman for
useful discussions and  critical comments, and the Centre for
Quantum Technologies of the National University of Singapore for hospitality.

\begin{widetext}
\appendix
\section{Real scalar field}\label{appendixa}

We adapt the following normalization convention for the sates and operators:
\be
|\bp\9=\alpha^\dag_\bp|0\9, \qquad  [\alpha_\bq,\alpha^\dag_\bp]=(2 \pi)^3 (2E_\bp) \delta^{(3)}(\bq-\bp)
\ee
so the scalar field is written as
\be
\hphi=\int\!d\mu(p)\left(\alpha_\bp e^{-ip\cdot x}+\alpha_\bp^\dag e^{ip\cdot x}\right).
\ee
For a one-particle state of Eq.~\eqref{scapsi} its wave function in the configuration  space is defined by a
generalized Fourier transform   as
  \be
  \psi(\bx,t)=\int d\mu(p)\psi(\bp)e^{i(\bp\cdot\bx-Et)}.
  \ee
The normal-ordered energy density  is
\begin{align}
 \hat{T}_{00}(t,\bx)& =::\!\!\cH(t,\bx)\!\!::=
 \tfrac{1}{2}\int\!d\mu(p)d\mu(q)\left(E_\bp E_\bq+\bp\cdot\bq+m^2\right)
\left(\alpha^\dag_\bp\alpha_\bq e^{i(p-q)\cdot x}+\alpha^\dag_\bq\alpha_\bp e^{-i(p-q)\cdot x}\right)+\ldots\nonumber \\
& =\int\!d\mu(p)d\mu(q)\left(E_\bp E_\bq+\bp\cdot\bq+m^2\right)\alpha^\dag_\bp\alpha_\bq e^{i(p-q)\cdot x},
\end{align}
where $\ldots$ terms have zero matrix elements between the states with the same definite number of particles.
Similarly, the normal-ordered momentum density is
\begin{align}
 \hat{T}_{0k}(t,\bx) = ::\!\!\frac{\pad \hphi}{\pad x^0}\frac{\pad \hphi}{\pad x^k}::=
  \int\!d\mu(p)d\mu(q)E_\bp q_k \left(\alpha^\dag_\bp\alpha_\bq e^{i(p-q)\cdot x}+
 \alpha^\dag_\bq\alpha_\bp e^{-i(p-q)\cdot x}\right)+\ldots
\end{align}

  The expectation values are
\begin{align}
\6\Psi|\hat{T}_{00}|\Psi\9=\int d\mu(k)d\mu(k')(E_\bk E_{\bk'}+\bk\cdot\bk'+m^2)\psi^*(\bk)\psi(\bk')e^{i(k-k')\cdot x} \nonumber \\
=|\nabla\psi(t,\bx)|^2+|\pad_t\psi(t,\bx)|^2+m^2|\psi(t,\bx)|^2,
\end{align}
and
\be
\6\Psi|\hat{T}_{0j}|\Psi\9=-2\int d\mu(k)d\mu(k')E_\bk k_j\psi^*(\bk)\psi(\bk')e^{i(k-k')\cdot x},
\ee
respectively.
Hence the Hamiltonian and its inverse square root are given by
\be
\hat{H}=\int\hT_{00}dV=\int d\mu(p)E_\bp \alpha^\dag_\bp\alpha_\bp, \qquad \hat{H}^{-1/2}=\int d\mu(p), \frac{1}{\sqrt{E_\bp}}\alpha^\dag_\bp\alpha_\bp
\ee
respectively, and the  POVM element $\hPi$ is given by
\be
\hPi(t,\bx)=\hat{H}^{-1/2}\hT_{00}(t,\bx)\hat{H}^{-1/2}=\int\!d\mu(p)d\mu(q)\frac{1}
{\sqrt{E_\bp E_\bq}}\left(E_\bp E_\bq+\bp\cdot\bq+m^2\right)\alpha^\dag_\bp\alpha_\bq e^{i(p-q)\cdot x}.
\ee
This results in the following probability density:
 \begin{align}
p(t,\bx)&=\6\Psi|\hPi(t,\bx)|\Psi\9=\int d\mu(k)d\mu(k')\frac{1}{\sqrt{E_\bk E_{\bk'}}}(E_\bk E_{\bk'}+\bk\cdot\bk'+m^2)
\psi^*(\bk)\psi(\bk')e^{i(k-k')\cdot x} \nonumber \\
&=|\nabla\tilde{\psi}(t,\bx)|^2+|\pad_t\tilde{\psi}(t,\bx)|^2
  +m^2|\tilde{\psi}(t,\bx)|^2. \label{probab}
\end{align}
Similarly,  the probability current is
\be
\hat{\jmath}_l(t,x)=\hat{H}^{-1/2}\hT_{0l}(t,\bx)\hat{H}^{-1/2}.
\ee

Introducing
\be
f(\bp,\bq):=\frac{1}{4E_\bp^{\,3/2}E_\bq^{\,3/2}}(E_\bp E_\bq+\bp\cdot\bq+m^2),
\ee
we express the expectation of $\bx$ at time $t=0$ as
\be
\6\bx(0)\9=-i\int\!d^3\bx\frac{d^3\bp}{(2\pi)^3}\frac{d^3\bq}{(2\pi)^3}
e^{i(p-q)\cdot x}\nabla_\bp\left(f(\bp,\bq)\psi^*(\bp)\psi(\bq)\right),
\ee
and the second moment as
\be
\6x^2_l(0)\9=\int\!d^3\bx\frac{d^3\bp}{(2\pi)^3}\frac{d^3\bq}{(2\pi)^3}
e^{i(p-q)\cdot x}\pad_{p_l}   \pad_{q_l}\left(f(\bp,\bq)\psi^*(\bp)\psi(\bq)\right).%
\ee
These may be more conveniently expressed as
\be
\6\bx\9=i\int\! \frac{d^3\bp}{(2\pi)^3} \left(\frac{\bp}{4E_\bp^3}|\psi(\bp)|^2-\frac{1}{2E_\bp}\psi(\bp)
\nabla_\bp\psi(\bp)^*\right),
\ee
and
\be
\6x^2_l\9=\int\ \frac{d^3\bp}{(2\pi)^3} \left( \frac{ 2E_\bp^2-p_l^2}{8E_\bp^5}|\psi(\bp)|^2-
\frac{ p_l}{2E_\bp^3} \pad_{p_l}|\psi(\bp)|^2+
\frac{1}{2E_\bp}\big|\pad_{p_l}\psi(\bp)\big|^2\right).
\ee

For $t\neq 0$ we make the substitution
\be
 i\frac{\pad f}{\pad p_k}\rightarrow\left(-\frac{\pad E_\bp}{\pad p_k}t+ i\frac{\pad  }{\pad p_k}\right)f,
\ee
hence, e.g.,
\begin{align}
\6\bx(t)\9&=-i\int\!d^3\bx\frac{d^3\bp}{(2\pi)^3}
\frac{d^3\bq}{(2\pi)^3}e^{i(p-q)\cdot x}\nabla_\bp\left(f(\bp,\bq)\psi^*(\bp)\psi(\bq)\right)
+t\int\!\frac{d^3\bp}{(2\pi)^3}\frac{\bp}{E_\bp} f(\bp,\bp)\psi^*(\bp)\psi(\bp)\nonumber \\
&=\6\bx(t=0)\9+\6{\mathbf {v}}\9t.
\end{align}

To obtain the nonrelativistic identification of a wave-function we expand Eq.~\eqref{probab} in powers of $\bp/m$, and find that the leading term  agrees with the nonrelativistic value,
\be
\lim_{|\bp|\ll m} \psi(\bp)/\sqrt{2m}=\psi(\bp)^\mathrm{nonrel}.
\ee

\section{Electromagnetic field}

In this section we restore the factors $\hbar$ and $c$. The classical gauge-invariant Lagrangian is
\be
\mathcal{L}=\tfrac{1}{4}F_{\mu\nu}F^{\mu\nu},
\ee
where $F_{\mu\nu}=\pad_\mu A_\nu-\pad_\nu A_\mu$ is the electromagnetic tensor.
We quantize the canonical variables  $\bA$ (configuration variables) and $-\bE$ (momenta) in the radiation gauge.
The creation and annihilation operators satisfy
\be
[\alpha_{\bp\lambda},\alpha_{\bq\lambda'}^\dag]=(2 \pi)^3 (2E_\bp)\delta_{\lambda\lambda'} \delta^{(3)}(\bq-\bp).
\ee
Hence
\be
\hat{\bA}(t,\bx)=\frac{c}{\sqrt{\hbar}}\int\! d\mu(p)\sum_{\lambda=\pm 1}\left(\bep^\lambda_\bp\alpha_{\bp\lambda}e^{-ip\cdot x/\hbar}+
\bep^{\lambda*}_\bp\alpha_{\bp\lambda}^\dag e^{ip\cdot x/\hbar}\right),\qquad d\mu(p)=\frac{1}{(2\pi)^3}\frac{d^3\bp}{(2E_\bp)},
\ee
and the field commutation relations are 
\be
[\hA^m(t,\bx),\hE^n(t,\by)]=-i\hbar c \delta^{mn}\delta^{(3)}(\bx-\by), \qquad
[\hB_k(t,\bx),\hE_l(t,\by)]=i\hbar c  \,\varepsilon_{klm}\pad_{x^m}\delta^{(3)}(\bx-\by),
\ee
where $\varepsilon_{klm}$ is the totally antisymmetric symbol.

Energy density takes the form
\begin{align}
 \hT_{00}  =\frac{1}{2}\!::\hbF^\dag\!\cdot\!\hbF::
=\frac{2}{\hbar^3}\!\int\!d\mu(p)d\mu(q)E_\bp E_\bq\,
 \bep_\bp^*\!\cdot\!\bep_\bq\big(\alpha^\dag_{\bp+}\alpha_{\bq+} e^{i(p-q)\cdot x/\hbar}+\alpha^\dag_{\bq-}\alpha_{\bp-}
  e^{-i(p-q)\cdot x/\hbar}\big),
\end{align}
where the terms that annihilate  states with a fixed number of particles are dropped.

Similarly, momentum density is
\begin{align}
 \hT_{0k}/c  =\frac{1}{2i c}::\hbF^\dag\!\times\!\hbF::
=\frac{2}{\hbar^3 c}\int\!d\mu(p)d\mu(q)E_\bp E_\bq\, \bep_\bp^*\!\times\!\bep_\bq\big(\alpha^\dag_{\bp+}
\alpha_{\bq+} e^{i(p-q)\cdot x/\hbar}+\alpha^\dag_{\bq-}\alpha_{\bp-} e^{-i(p-q)\cdot x/\hbar}\big),
\end{align}
and  the Hamiltonian is
\be
\hH=\int\! d\mu(p)E_\bp\sum_{\lambda=\pm 1}\alpha^\dag_{\bp\lambda}\alpha_{\bp\lambda}.
\ee

The probability density $p(t,\bx)=\6\Psi\hPi(t,\bx)|\Psi\9$ equals
\be
p(t,\bx)=\frac{2 }{\hbar^3}\int\!d\mu(k)d\mu(k')\sqrt{E_\bk E_{\bk'}}\, \big(\bep_\bk^* \cdot \bep_{\bk'}
\psi^*_+(\bk)\psi_{+}(\bk')+\bep_\bk \cdot \bep_{\bk'}^*\psi^*_-(\bk)\psi_{-}(\bk')\big)e^{i(k-k')\cdot x/\hbar}.
\ee
Similarly, the probability current is given by
\be
\hat{\jmath}_m(t,\bx)=\6\Psi|\hat{H}^{-1/2}\hT_{0m}(t,\bx)\hat{H}^{-1/2}|\Psi\9/c.
\ee

\end{widetext}

  \end{document}